# Compounding Fields and Their Quantum Equations in the Sedenion Space

Zihua Weng

*School of Physics and Mechanical & Electrical Engineering, P. O. Box 310,*
*Xiamen University, Xiamen 361005, China*

**Abstract**

The sedenion compounding fields and their quantum interplays can be presented by analogy with the octonionic electromagnetic, gravitational, strong and weak interactions. In sedenion fields which are associated with electromagnetic, gravitational, strong and weak interactions, the study deduces some conclusions of field source particles and intermediate particles which are consistent with current electro-weak theory. In the sedenion fields which are associated with the hyper-strong and strong-weak fields, the paper draws some predicts and conclusions of the field source particles and intermediate particles. The researches results show that there may exist some new field source particles in the sedenion compounding fields.

## 1. Introduction

Nowadays, there still exist some movement phenomena which can't be explained by current electro-weak theory. Therefore, some scientists doubt the universality of current electro-weak theory, and then bring forward some new field theories to explain the abnormal phenomena in the electromagnetic interaction and weak interaction.

A novel insight on the problem of electromagnetic and weak interactions can be given by the concept of sedenion space. By analogy with the octonionic fields, some sorts of sedenion fields which are consist of octonionic fields S-W and E-G etc., can be obtained in the paper. The paper describes the sedenion fields and their quantum theory, and draws some predicts and new conclusions which are consistent with the current electro-weak theory etc.

## 2. Compounding fields in the sedenion spaces

Through the analysis of the different fields in octonionic spaces, we find that each interaction possesses its own spacetime, field and operator in accordance with the 'SpaceTime Equality Postulation'. In the octonionic spaces [1, 2], four sorts of octonionic fields can be tabulated as Table 1, including their operators, spaces and fields.

---

*E-mail Addresses*: xmuwzh@hotmail.com, xmuwzh@xmu.edu.cn

Table 1.  The compounding fields and operators in octonionic spaces

| operator | $\mathcal{X}_{H-S} / k^X_{H-S} + \diamond$ | $\mathcal{A}_{S-W} / k^A_{S-W} + \diamond$ | $\mathcal{B}_{E-G} / k^B_{E-G} + \diamond$ | $\mathcal{S}_{H-W} / k^S_{H-W} + \diamond$ |
|---|---|---|---|---|
| space | octonion space H-S | octonion space S-W | octonion space E-G | octonion space H-W |
| field | H-S | S-W | E-G | H-W |

By analogy with the cases in the different octonionic spaces, the operators and fields in the different sedenion spaces can be written as Table 2. There exist sixteen sorts of compounding fields in the sedenion spaces, including four sorts of octonionic fields. The study will show typically their four kinds of fields in the sedenion spaces.

Table 2.  The compounding fields and operators in the different sedenion spaces

| operator | $\mathcal{X}_{H-S} / k^X_{H-S} + \diamond$ | $\diamond + \mathcal{A}_{S-W} / k^A_{S-W} + \mathcal{A}_{H-S} / k^A_{H-S}$ | $\diamond + \mathcal{B}_{E-G} / k^B_{E-G} + \mathcal{B}_{H-S} / k^B_{H-S}$ | $\diamond + \mathcal{S}_{H-W} / k^S_{H-W} + \mathcal{S}_{H-S} / k^S_{H-S}$ |
|---|---|---|---|---|
| space | octonion space H-S | sedenion space SW-HS | sedenion space EG-HS | sedenion space HW-HS |
| field | H-S | SW-HS | EG-HS | HW-HS |
| operator | $\diamond + \mathcal{X}_{H-S} / k^X_{H-S} + \mathcal{X}_{S-W} / k^X_{S-W}$ | $\mathcal{A}_{S-W} / k^A_{S-W} + \diamond$ | $\diamond + \mathcal{B}_{E-G} / k^B_{E-G} + \mathcal{B}_{S-W} / k^B_{S-W}$ | $\diamond + \mathcal{S}_{H-W} / k^S_{H-W} + \mathcal{S}_{S-W} / k^S_{S-W}$ |
| space | sedenion space HS-SW | octonion space S-W | sedenion space EG-SW | sedenion space HW-SW |
| field | HS-SW | S-W | EG-SW | HW-SW |
| operator | $\diamond + \mathcal{X}_{H-S} / k^X_{H-S} + \mathcal{X}_{E-G} / k^X_{E-G}$ | $\diamond + \mathcal{A}_{S-W} / k^A_{S-W} + \mathcal{A}_{E-G} / k^A_{E-G}$ | $\mathcal{B}_{E-G} / k^B_{E-G} + \diamond$ | $\diamond + \mathcal{S}_{H-W} / k^S_{H-W} + \mathcal{S}_{E-G} / k^S_{E-G}$ |
| space | sedenion space HS-EG | sedenion space SW-EG | octonion space E-G | sedenion space HW-EG |
| field | HS-EG | SW-EG | E-G | HW-EG |
| operator | $\diamond + \mathcal{X}_{H-S} / k^X_{H-S} + \mathcal{X}_{H-W} / k^X_{H-W}$ | $\diamond + \mathcal{A}_{S-W} / k^A_{S-W} + \mathcal{A}_{H-W} / k^A_{H-W}$ | $\diamond + \mathcal{B}_{E-G} / k^B_{E-G} + \mathcal{B}_{H-W} / k^B_{H-W}$ | $\mathcal{S}_{H-W} / k^S_{H-W} + \diamond$ |
| space | sedenion space HS-HW | sedenion space SW-HW | sedenion space EG-HW | octonion space H-W |
| field | HS-HW | SW-HW | EG-HW | H-W |

## 3. Compounding field in sedenion space SW-EG

It is believed that strong-weak field and electromagnetic-gravitational field are unified, equal and interconnected. By means of the conception of the space expansion etc., two types of the octonionic spaces (Octonionic space S-W and Octonionic space E-G) can be combined into a sedenion space (Sedenion space SW-EG, for short). In the sedenion space, some properties of strong, weak, electromagnetic and gravitational interactions can be described uniformly.

In the sedenion space SW-EG, there exist one kind of field (sedenion field SW-EG, for short) can be obtained related to the operator ($\mathcal{A}/k + \diamondsuit$).

Table 3. Sedenion multiplication table

| × | 1 | $e_1$ | $e_2$ | $e_3$ | $E_4$ | $e_5$ | $e_6$ | $e_7$ | $e_8$ | $e_9$ | $e_{10}$ | $e_{11}$ | $e_{12}$ | $e_{13}$ | $e_{14}$ | $e_{15}$ |
|---|---|---|---|---|---|---|---|---|---|---|---|---|---|---|---|---|
| 1 | 1 | $e_1$ | $e_2$ | $e_3$ | $E_4$ | $e_5$ | $e_6$ | $e_7$ | $e_8$ | $e_9$ | $e_{10}$ | $e_{11}$ | $e_{12}$ | $e_{13}$ | $e_{14}$ | $e_{15}$ |
| $e_1$ | $e_1$ | -1 | $e_3$ | $-e_2$ | $E_5$ | $-e_4$ | $-e_7$ | $e_6$ | $e_9$ | $-e_8$ | $-e_{11}$ | $e_{10}$ | $-e_{13}$ | $e_{12}$ | $e_{15}$ | $-e_{14}$ |
| $e_2$ | $e_2$ | $-e_3$ | -1 | $e_1$ | $E_6$ | $e_7$ | $-e_4$ | $-e_5$ | $e_{10}$ | $e_{11}$ | $-e_8$ | $-e_9$ | $-e_{14}$ | $-e_{15}$ | $e_{12}$ | $e_{13}$ |
| $e_3$ | $e_3$ | $e_2$ | $-e_1$ | -1 | $E_7$ | $-e_6$ | $e_5$ | $-e_4$ | $e_{11}$ | $-e_{10}$ | $e_9$ | $-e_8$ | $-e_{15}$ | $e_{14}$ | $-e_{13}$ | $e_{12}$ |
| $e_4$ | $e_4$ | $-e_5$ | $-e_6$ | $-e_7$ | -1 | $e_1$ | $e_2$ | $e_3$ | $e_{12}$ | $e_{13}$ | $e_{14}$ | $e_{15}$ | $-e_8$ | $-e_9$ | $-e_{10}$ | $-e_{11}$ |
| $e_5$ | $e_5$ | $e_4$ | $-e_7$ | $e_6$ | $-e_1$ | -1 | $-e_3$ | $e_2$ | $e_{13}$ | $-e_{12}$ | $e_{15}$ | $-e_{14}$ | $e_9$ | $-e_8$ | $e_{11}$ | $-e_{10}$ |
| $e_6$ | $e_6$ | $e_7$ | $e_4$ | $-e_5$ | $-e_2$ | $e_3$ | -1 | $-e_1$ | $e_{14}$ | $-e_{15}$ | $-e_{12}$ | $e_{13}$ | $e_{10}$ | $-e_{11}$ | $-e_8$ | $e_9$ |
| $e_7$ | $e_7$ | $-e_6$ | $e_5$ | $e_4$ | $-e_3$ | $-e_2$ | $e_1$ | -1 | $e_{15}$ | $e_{14}$ | $-e_{13}$ | $-e_{12}$ | $e_{11}$ | $e_{10}$ | $-e_9$ | $-e_8$ |
| $e_8$ | $e_8$ | $-e_9$ | $-e_{10}$ | $-e_{11}$ | $-e_{12}$ | $-e_{13}$ | $-e_{14}$ | $-e_{15}$ | -1 | $e_1$ | $e_2$ | $e_3$ | $e_4$ | $e_5$ | $e_6$ | $e_7$ |
| $e_9$ | $e_9$ | $e_8$ | $-e_{11}$ | $e_{10}$ | $-e_{13}$ | $e_{12}$ | $e_{15}$ | $-e_{14}$ | $-e_1$ | -1 | $-e_3$ | $e_2$ | $-e_5$ | $e_4$ | $e_7$ | $-e_6$ |
| $e_{10}$ | $e_{10}$ | $e_{11}$ | $e_8$ | $-e_9$ | $-e_{14}$ | $-e_{15}$ | $e_{12}$ | $e_{13}$ | $-e_2$ | $e_3$ | -1 | $-e_1$ | $-e_6$ | $-e_7$ | $e_4$ | $e_5$ |
| $e_{11}$ | $e_{11}$ | $-e_{10}$ | $e_9$ | $e_8$ | $-e_{15}$ | $e_{14}$ | $-e_{13}$ | $e_{12}$ | $-e_3$ | $-e_2$ | $e_1$ | -1 | $-e_7$ | $e_6$ | $-e_5$ | $e_4$ |
| $e_{12}$ | $e_{12}$ | $e_{13}$ | $e_{14}$ | $e_{15}$ | $E_8$ | $-e_9$ | $-e_{10}$ | $-e_{11}$ | $-e_4$ | $e_5$ | $e_6$ | $e_7$ | -1 | $-e_1$ | $-e_2$ | $-e_3$ |
| $e_{13}$ | $e_{13}$ | $-e_{12}$ | $e_{15}$ | $-e_{14}$ | $E_9$ | $e_8$ | $e_{11}$ | $-e_{10}$ | $-e_5$ | $-e_4$ | $e_7$ | $-e_6$ | $e_1$ | -1 | $e_3$ | $-e_2$ |
| $e_{14}$ | $e_{14}$ | $-e_{15}$ | $-e_{12}$ | $e_{13}$ | $E_{10}$ | $-e_{11}$ | $e_8$ | $e_9$ | $-e_6$ | $-e_7$ | $-e_4$ | $e_5$ | $e_2$ | $-e_3$ | -1 | $e_1$ |
| $e_{15}$ | $e_{15}$ | $e_{14}$ | $-e_{13}$ | $-e_{12}$ | $E_{11}$ | $e_{10}$ | $-e_9$ | $e_8$ | $-e_7$ | $e_6$ | $-e_5$ | $-e_4$ | $e_3$ | $e_2$ | $-e_1$ | -1 |

In the octonionic space S-W, the base $\mathcal{E}_{S-W}$ can be written as

$$\mathcal{E}_{S-W} = (1, \vec{e}_1, \vec{e}_2, \vec{e}_3, \vec{e}_4, \vec{e}_5, \vec{e}_6, \vec{e}_7) \tag{1}$$

The displacement ($r_0, r_1, r_2, r_3, r_4, r_5, r_6, r_7$) in the octonionic space S-W is

$$\mathcal{R}_{S-W} = r_0 + \vec{e}_1 r_1 + \vec{e}_2 r_2 + \vec{e}_3 r_3 + \vec{e}_4 r_4 + \vec{e}_5 r_5 + \vec{e}_6 r_6 + \vec{e}_7 r_7 \tag{2}$$

where, $r_0 = c_{S-W} t_{S-W}$, $r_4 = c_{S-W} T_{S-W}$. $c_{S-W}$ is the speed of intermediate particle in the strong-weak field, $t_{S-W}$ and $T_{S-W}$ denote the time.

The octonionic differential operator $\diamondsuit_{SWEG1}$ and its conjugate operator are defined as

$$\diamondsuit_{SWEG1} = \partial_0 + \vec{e}_1 \partial_1 + \vec{e}_2 \partial_2 + \vec{e}_3 \partial_3 + \vec{e}_4 \partial_4 + \vec{e}_5 \partial_5 + \vec{e}_6 \partial_6 + \vec{e}_7 \partial_7 \tag{3}$$

$$\diamondsuit^*_{SWEG1} = \partial_0 - \vec{e}_1 \partial_1 - \vec{e}_2 \partial_2 - \vec{e}_3 \partial_3 - \vec{e}_4 \partial_4 - \vec{e}_5 \partial_5 - \vec{e}_6 \partial_6 - \vec{e}_7 \partial_7 \tag{4}$$

where, $\partial_j = \partial / \partial r_j$, $j = 0, 1, 2, 3, 4, 5, 6, 7$.

In the octonionic space E-G, the base $\mathcal{E}_{E-G}$ can be written as

$$\mathcal{E}_{E-G} = (\vec{e}_8, \vec{e}_9, \vec{e}_{10}, \vec{e}_{11}, \vec{e}_{12}, \vec{e}_{13}, \vec{e}_{14}, \vec{e}_{15}) \tag{5}$$

The displacement ($r_8, r_9, r_{10}, r_{11}, r_{12}, r_{13}, r_{14}, r_{15}$) in the octonionic space E-G is

$$\mathcal{R}_{E-G} = \vec{e}_8 r_8 + \vec{e}_9 r_9 + \vec{e}_{10} r_{10} + \vec{e}_{11} r_{11} + \vec{e}_{12} r_{12} + \vec{e}_{13} r_{13} + \vec{e}_{14} r_{14} + \vec{e}_{15} r_{15} \tag{6}$$

where, $r_8 = c_{E-G} t_{E-G}$, $r_{12} = c_{E-G} T_{E-G}$. $c_{E-G}$ is the speed of intermediate particle in the electromagnetic-gravitational field, $t_{E-G}$ and $T_{E-G}$ denote the time.

The octonionic differential operator $\diamondsuit_{SWEG2}$ and its conjugate operator are defined as,

$$\diamondsuit_{SWEG2} = \vec{e}_8 \partial_8 + \vec{e}_9 \partial_9 + \vec{e}_{10} \partial_{10} + \vec{e}_{11} \partial_{11}$$

$$+ \vec{e}_{12} \partial_{12} + \vec{e}_{13} \partial_{13} + \vec{e}_{14} \partial_{14} + \vec{e}_{15} \partial_{15} \qquad (7)$$

$$\lozenge^*_{SWEG2} = -\vec{e}_8 \partial_8 - \vec{e}_9 \partial_9 - \vec{e}_{10} \partial_{10} - \vec{e}_{11} \partial_{11}$$
$$- \vec{e}_{12} \partial_{12} - \vec{e}_{13} \partial_{13} - \vec{e}_{14} \partial_{14} - \vec{e}_{15} \partial_{15} \qquad (8)$$

where, $\partial_j = \partial/\partial r_j$, $j = 8, 9, 10, 11, 12, 13, 14, 15$.

In the sedenion space SW-EG, the base $\mathcal{E}_{SW-EG}$ can be written as

$$\mathcal{E}_{SW-EG} = \mathcal{E}_{S-W} + \mathcal{E}_{E-G}$$
$$= (1, \vec{e}_1, \vec{e}_2, \vec{e}_3, \vec{e}_4, \vec{e}_5, \vec{e}_6, \vec{e}_7, \vec{e}_8, \vec{e}_9, \vec{e}_{10}, \vec{e}_{11}, \vec{e}_{12}, \vec{e}_{13}, \vec{e}_{14}, \vec{e}_{15}) \qquad (9)$$

The displacement ($r_0$, $r_1$, $r_2$, $r_3$, $r_4$, $r_5$, $r_6$, $r_7$, $r_8$, $r_9$, $r_{10}$, $r_{11}$, $r_{12}$, $r_{13}$, $r_{14}$, $r_{15}$) in the sedenion space SW-EG is

$$\mathcal{R}_{SW-EG} = \mathcal{R}_{S-W} + \mathcal{R}_{E-G}$$
$$= r_0 + \vec{e}_1 r_1 + \vec{e}_2 r_2 + \vec{e}_3 r_3 + \vec{e}_4 r_4 + \vec{e}_5 r_5 + \vec{e}_6 r_6 + \vec{e}_7 r_7 + \vec{e}_8 r_8$$
$$+ \vec{e}_9 r_9 + \vec{e}_{10} r_{10} + \vec{e}_{11} r_{11} + \vec{e}_{12} r_{12} + \vec{e}_{13} r_{13} + \vec{e}_{14} r_{14} + \vec{e}_{15} r_{15} \qquad (10)$$

The sedenion differential operator $\lozenge_{SWEG}$ and its conjugate operator are defined as

$$\lozenge_{SWEG} = \lozenge_{SWEG1} + \lozenge_{SWEG2} \qquad (11)$$

$$\lozenge^*_{SWEG} = \lozenge^*_{SWEG1} + \lozenge^*_{SWEG2} \qquad (12)$$

Table 4. Subfields of sedenion field SW-EG

|  | Strong Interaction | Weak Interaction | Electromagnetic Interaction | Gravitational Interaction |
|---|---|---|---|---|
| strong quaternionic space S | strong-strong subfield S strong-charge $\gamma^B_{ss}$ | weak-strong subfield S weak-charge $\gamma^B_{ws}$ | electromagnetic-strong subfield S electromagnetic-charge, $\gamma^B_{es}$ | gravitational-strong subfield S gravitational-charge, $\gamma^B_{gs}$ |
| weak quaternionic space W | strong-weak subfield W strong-charge $\gamma^B_{sw}$ | weak-weak subfield W weak-charge $\gamma^B_{ww}$ | electromagnetic-weak subfield W electromagnetic-charge, $\gamma^B_{ew}$ | gravitational-weak subfield, W gravitational-charge, $\gamma^B_{gw}$ |
| electromagnetic quaternionic space E | strong-electromagnetic subfield E strong-charge, $\gamma^B_{se}$ | weak-electromagnetic subfield E weak-charge, $\gamma^B_{we}$ | electromagnetic-electromagnetic subfield, E electromagnetic-charge, $\gamma^B_{ee}$ | gravitational-electromagnetic subfield, E gravitational-charge, $\gamma^B_{ge}$ |
| gravitational quaternionic space G | strong-gravitational subfield G strong-charge, $\gamma^B_{sg}$ | weak-gravitational subfield G weak-charge, $\gamma^B_{wg}$ | electromagnetic-gravitational subfield, G electromagnetic-charge, $\gamma^B_{eg}$ | gravitational-gravitational subfield, G gravitational-charge, $\gamma^B_{gg}$ |

In the sedenion field SW-EG, the definition of the field strength is very different to that of the octonionic field, and thus the succeeding equations and operators should be revised. By analogy with the case of other fields, the sedenion differential operator $\Diamond$ needs to be generalized to the new operator ($\mathcal{A}_{S\text{-}W} / k^A_{S\text{-}W} + \mathcal{A}_{E\text{-}G} / k^A_{E\text{-}G} + \Diamond$). This is because the sedenion field SW-EG includes strong-weak field and electromagnetic-gravitational field.

It can be predicted that the field strength of the off-diagonal subfields may be symmetry along the diagonal subfield in the Table 4. For example, the field strength of the weak-electromagnetic subfield may be equal to that of the electromagnetic-weak subfield. And it is easy to find that the weak interaction is interconnected with electromagnetic interaction [3]. The physical features of each subfield in the sedenion field SW-EG meet the requirement of the equations set in the Table 5.

In the sedenion field SW-EG, the field potential ($a_0$, $a_1$, $a_2$, $a_3$, $a_4$, $a_5$, $a_6$, $a_7$, $a_8$, $a_9$, $a_{10}$, $a_{11}$, $a_{12}$, $a_{13}$, $a_{14}$, $a_{15}$) is defined as

$$\mathcal{A} = \Diamond^* \circ \mathcal{X}$$
$$= a_0 + a_1 \vec{e}_1 + a_2 \vec{e}_2 + a_3 \vec{e}_3 + a_4 \vec{e}_4 + a_5 \vec{e}_5$$
$$+ a_6 \vec{e}_6 + a_7 \vec{e}_7 + a_8 \vec{e}_8 + a_9 \vec{e}_9 + a_{10} \vec{e}_{10}$$
$$+ a_{11} \vec{e}_{11} + a_{12} \vec{e}_{12} + a_{13} \vec{e}_{13} + a_{14} \vec{e}_{14} + a_{15} \vec{e}_{15} \qquad (13)$$

where, $k_{rx} \mathcal{X} = k^s_{rx} \mathcal{X}_s + k^w_{rx} \mathcal{X}_w + k^e_{rx} \mathcal{X}_e + k^g_{rx} \mathcal{X}_g$. $\mathcal{X}_s$ is the physical quantity in S space, $\mathcal{X}_w$ is the physical quantity in W space, $\mathcal{X}_e$ is the physical quantity in E space, and $\mathcal{X}_g$ is the physical quantity in G space. $k_{rx}$, $k^s_{rx}$, $k^w_{rx}$, $k^e_{rx}$ and $k^g_{rx}$ are coefficients.

The field strength $\mathcal{B}$ of the sedenion field SW-EG can be defined as

$$\mathcal{B} = (\mathcal{A}/K + \Diamond) \circ \mathcal{A} = (\mathcal{A}_{S\text{-}W} / k^A_{S\text{-}W} + \mathcal{A}_{E\text{-}G} / k^A_{E\text{-}G} + \Diamond) \circ \mathcal{A} \qquad (14)$$

where, $k^A_{S\text{-}W}$, $k^A_{E\text{-}G}$ and $K$ are coefficients. The field potential $\mathcal{A}_{S\text{-}W}$ of the strong-weak field and the field potential $\mathcal{A}_{E\text{-}G}$ of the electromagnetic-gravitational field are

$$\mathcal{A}_{S\text{-}W} = a_0 + a_1 \vec{e}_1 + a_2 \vec{e}_2 + a_3 \vec{e}_3 + a_4 \vec{e}_4 + a_5 \vec{e}_5 + a_6 \vec{e}_6 + a_7 \vec{e}_7$$
$$\mathcal{A}_{E\text{-}G} = a_8 \vec{e}_8 + a_9 \vec{e}_9 + a_{10} \vec{e}_{10} + a_{11} \vec{e}_{11} + a_{12} \vec{e}_{12} + a_{13} \vec{e}_{13} + a_{14} \vec{e}_{14} + a_{15} \vec{e}_{15}$$

Then the field source and the force of the sedenion field can be defined respectively as

$$K \mu \mathcal{S} = K (\mathcal{A}_{S\text{-}W} / k^A_{S\text{-}W} + \mathcal{A}_{E\text{-}G} / k^A_{E\text{-}G} + \Diamond)^* \circ \mathcal{B} \qquad (15)$$
$$\mathcal{Z} = K (\mathcal{A}_{S\text{-}W} / k^A_{S\text{-}W} + \mathcal{A}_{E\text{-}G} / k^A_{E\text{-}G} + \Diamond) \circ \mathcal{S} \qquad (16)$$

where, the mark (*) denotes sedenion conjugate. The coefficient $\mu$ is interaction intensity of the sedenion field SW-EG. $K = K^A_{SWEG}$ is the coefficient.

Table 5. Equations set of sedenion field SW-EG

| Spacetime | Sedenion space SW-EG |
| --- | --- |
| $\mathcal{X}$ physical quantity | $k_{rx} \mathcal{X} = k^s_{rx} \mathcal{X}_s + k^w_{rx} \mathcal{X}_w + k^e_{rx} \mathcal{X}_e + k^g_{rx} \mathcal{X}_g$ |
| Field potential | $\mathcal{A} = \Diamond^* \circ \mathcal{X}$ |
| Field strength | $\mathcal{B} = (\mathcal{A}_{S\text{-}W} / k^A_{S\text{-}W} + \mathcal{A}_{E\text{-}G} / k^A_{E\text{-}G} + \Diamond) \circ \mathcal{A}$ |
| Field source | $K \mu \mathcal{S} = K (\mathcal{A}_{S\text{-}W} / k^A_{S\text{-}W} + \mathcal{A}_{E\text{-}G} / k^A_{E\text{-}G} + \Diamond)^* \circ \mathcal{B}$ |
| Force | $\mathcal{Z} = K (\mathcal{A}_{S\text{-}W} / k^A_{S\text{-}W} + \mathcal{A}_{E\text{-}G} / k^A_{E\text{-}G} + \Diamond) \circ \mathcal{S}$ |
| Angular momentum | $\mathcal{M} = \mathcal{S} \circ (\mathcal{R} + k_{rx} \mathcal{X})$ |
| Energy | $\mathcal{W} = K (\mathcal{A}_{S\text{-}W} / k^A_{S\text{-}W} + \mathcal{A}_{E\text{-}G} / k^A_{E\text{-}G} + \Diamond)^* \circ \mathcal{M}$ |
| Power | $\mathcal{N} = K (\mathcal{A}_{S\text{-}W} / k^A_{S\text{-}W} + \mathcal{A}_{E\text{-}G} / k^A_{E\text{-}G} + \Diamond) \circ \mathcal{W}$ |

The angular momentum of the sedenion field can be defined as

$$\mathcal{M} = \mathcal{S} \circ (\mathcal{R} + k_{rx}\mathcal{X}) \qquad (17)$$

and the energy and power in the sedenion field can be defined respectively as

$$\mathcal{W} = K(\mathcal{A}_{S-W}/k^A{}_{S-W} + \mathcal{A}_{E-G}/k^A{}_{E-G} + \diamond)^* \circ \mathcal{M} \qquad (18)$$

$$\mathcal{N} = K(\mathcal{A}_{S-W}/k^A{}_{S-W} + \mathcal{A}_{E-G}/k^A{}_{E-G} + \diamond) \circ \mathcal{W} \qquad (19)$$

In the sedenion space SW-EG, the wave functions of quantum mechanics are the sedenion equations set. The Dirac and Klein-Gordon equations of the quantum mechanics are actually the wave equations set which are associated with the particle's angular momentum.

In sedenion field SW-EG, the Dirac equation and Klein-Gordon equation can be attained respectively from the energy equation (18) and power equation (19) after substituting operator $K(\mathcal{A}_{S-W}/k^A{}_{S-W} + \mathcal{A}_{E-G}/k^A{}_{E-G} + \diamond)$ for $(\mathcal{W}_{S-W}/k^A{}_{S-W}b^A{}_{S-W} + \mathcal{W}_{E-G}/k^A{}_{E-G}b^A{}_{E-G} + \diamond)$. The coefficients $b^A{}_{S-W}$ and $b^A{}_{E-G}$ are the Plank-like constant.

The $\mathcal{U}$ equation of the quantum mechanics can be defined as

$$\mathcal{U} = (\mathcal{W}_{S-W}/k^A{}_{S-W}b^A{}_{S-W} + \mathcal{W}_{E-G}/k^A{}_{E-G}b^A{}_{E-G} + \diamond)^* \circ \mathcal{M} \qquad (20)$$

The $\mathcal{L}$ equation of the quantum mechanics can be defined as

$$\mathcal{L} = (\mathcal{W}_{S-W}/k^A{}_{S-W}b^A{}_{S-W} + \mathcal{W}_{E-G}/k^A{}_{E-G}b^A{}_{E-G} + \diamond) \circ \mathcal{U} \qquad (21)$$

The three sorts of Dirac-like equations can be obtained from the Eqs.(14), (15) and (16) respectively.

The $\mathcal{G}$ equation of the quantum mechanics can be defined as

$$\mathcal{G} = (\mathcal{W}_{S-W}/k^A{}_{S-W}b^A{}_{S-W} + \mathcal{W}_{E-G}/k^A{}_{E-G}b^A{}_{E-G} + \diamond) \circ \mathcal{A} \qquad (22)$$

The $\mathcal{T}$ equation of the quantum mechanics can be defined as

$$\mathcal{T} = (\mathcal{W}_{S-W}/k^A{}_{S-W}b^A{}_{S-W} + \mathcal{W}_{E-G}/k^A{}_{E-G}b^A{}_{E-G} + \diamond)^* \circ \mathcal{G} \qquad (23)$$

The $O$ equation of the quantum mechanics can be defined as

$$O = (\mathcal{W}_{S-W}/k^A{}_{S-W}b^A{}_{S-W} + \mathcal{W}_{E-G}/k^A{}_{E-G}b^A{}_{E-G} + \diamond) \circ \mathcal{T} \qquad (24)$$

Table 6.  Quantum equations set of sedenion field SW-EG

| | |
|---|---|
| Energy quantum | $\mathcal{U} = (\mathcal{W}_{S-W}/k^A{}_{S-W}b^A{}_{S-W} + \mathcal{W}_{E-G}/k^A{}_{E-G}b^A{}_{E-G} + \diamond)^* \circ \mathcal{M}$ |
| Power quantum | $\mathcal{L} = (\mathcal{W}_{S-W}/k^A{}_{S-W}b^A{}_{S-W} + \mathcal{W}_{E-G}/k^A{}_{E-G}b^A{}_{E-G} + \diamond) \circ \mathcal{U}$ |
| Field strength quantum | $\mathcal{G} = (\mathcal{W}_{S-W}/k^A{}_{S-W}b^A{}_{S-W} + \mathcal{W}_{E-G}/k^A{}_{E-G}b^A{}_{E-G} + \diamond) \circ \mathcal{A}$ |
| Field source quantum | $\mathcal{T} = (\mathcal{W}_{S-W}/k^A{}_{S-W}b^A{}_{S-W} + \mathcal{W}_{E-G}/k^A{}_{E-G}b^A{}_{E-G} + \diamond)^* \circ \mathcal{G}$ |
| Force quantum | $O = (\mathcal{W}_{S-W}/k^A{}_{S-W}b^A{}_{S-W} + \mathcal{W}_{E-G}/k^A{}_{E-G}b^A{}_{E-G} + \diamond) \circ \mathcal{T}$ |

In the sedenion field SW-EG, the intermediate and field source particles can be obtained. We can find the intermediate particles and other kinds of new and unknown particles which may be existed in the nature.

## 4. Compounding field in sedenion space EG-SW

It is believed that electromagnetic-gravitational field and strong-weak field are unified, equal and interconnected [4, 5]. By means of the conception of the space expansion etc., two types of the octonionic spaces (Octonionic space E-G and Octonionic space S-W) can be combined into a sedenion space (Sedenion space EG-SW, for short), which is different to Sedenion space

SW-EG. In the sedenion space EG-SW, some properties of strong, weak, electromagnetic and gravitational interactions can be described uniformly.

In sedenion space EG-SW, there exist a sedenion field (sedenion field EG-SW, for short) which is different to sedenion field SW-EG, can be obtained related to the operator ($\mathcal{B}/k + \diamondsuit$).

In the octonionic space E-G, the base $\mathcal{E}_{E-G}$ can be written as

$$\mathcal{E}_{E-G} = (\mathbf{1},\ \vec{e}_1,\ \vec{e}_2,\ \vec{e}_3,\ \vec{e}_4,\ \vec{e}_5,\ \vec{e}_6,\ \vec{e}_7) \tag{25}$$

The displacement ($r_0, r_1, r_2, r_3, r_4, r_5, r_6, r_7$) in the octonionic space E-G is

$$\mathcal{R}_{E-G} = r_0 + \vec{e}_1 r_1 + \vec{e}_2 r_2 + \vec{e}_3 r_3 + \vec{e}_4 r_4 + \vec{e}_5 r_5 + \vec{e}_6 r_6 + \vec{e}_7 r_7 \tag{26}$$

where, $r_0 = c_{E-G} t_{E-G}$, $r_4 = c_{E-G} T_{E-G}$. $c_{E-G}$ is the speed of intermediate particle in the electromagnetic-gravitational field, $t_{E-G}$ and $T_{E-G}$ denote the time.

The octonionic differential operator $\diamondsuit_{EGSW1}$ and its conjugate operator are defined as

$$\diamondsuit_{EGSW1} = \partial_0 + \vec{e}_1 \partial_1 + \vec{e}_2 \partial_2 + \vec{e}_3 \partial_3 + \vec{e}_4 \partial_4 + \vec{e}_5 \partial_5 + \vec{e}_6 \partial_6 + \vec{e}_7 \partial_7 \tag{27}$$

$$\diamondsuit^*_{EGSW1} = \partial_0 - \vec{e}_1 \partial_1 - \vec{e}_2 \partial_2 - \vec{e}_3 \partial_3 - \vec{e}_4 \partial_4 - \vec{e}_5 \partial_5 - \vec{e}_6 \partial_6 - \vec{e}_7 \partial_7 \tag{28}$$

where, $\partial_j = \partial / \partial r_j$, $j = 0, 1, 2, 3, 4, 5, 6, 7$.

In the octonionic space S-W, the base $\mathcal{E}_{S-W}$ can be written as

$$\mathcal{E}_{S-W} = (\vec{e}_8,\ \vec{e}_9,\ \vec{e}_{10},\ \vec{e}_{11},\ \vec{e}_{12},\ \vec{e}_{13},\ \vec{e}_{14},\ \vec{e}_{15}) \tag{29}$$

The displacement ($r_8, r_9, r_{10}, r_{11}, r_{12}, r_{13}, r_{14}, r_{15}$) in the octonionic space S-W is

$$\mathcal{R}_{S-W} = \vec{e}_8 r_8 + \vec{e}_9 r_9 + \vec{e}_{10} r_{10} + \vec{e}_{11} r_{11} + \vec{e}_{12} r_{12} + \vec{e}_{13} r_{13} + \vec{e}_{14} r_{14} + \vec{e}_{15} r_{15} \tag{30}$$

where, $r_8 = c_{S-W} t_{S-W}$, $r_{12} = c_{S-W} T_{S-W}$. $c_{S-W}$ is the speed of intermediate particle in the strong-weak field, $t_{S-W}$ and $T_{S-W}$ denote the time.

The octonionic differential operator $\diamondsuit_{EGSW2}$ and its conjugate operator are defined as,

$$\begin{aligned}\diamondsuit_{EGSW2} = &\ \vec{e}_8 \partial_8 + \vec{e}_9 \partial_9 + \vec{e}_{10} \partial_{10} + \vec{e}_{11} \partial_{11} \\ &+ \vec{e}_{12} \partial_{12} + \vec{e}_{13} \partial_{13} + \vec{e}_{14} \partial_{14} + \vec{e}_{15} \partial_{15}\end{aligned} \tag{31}$$

$$\begin{aligned}\diamondsuit^*_{EGSW2} = &\ -\vec{e}_8 \partial_8 - \vec{e}_9 \partial_9 - \vec{e}_{10} \partial_{10} - \vec{e}_{11} \partial_{11} \\ &- \vec{e}_{12} \partial_{12} - \vec{e}_{13} \partial_{13} - \vec{e}_{14} \partial_{14} - \vec{e}_{15} \partial_{15}\end{aligned} \tag{32}$$

where, $\partial_j = \partial / \partial r_j$, $j = 8, 9, 10, 11, 12, 13, 14, 15$.

In the sedenion space EG-SW, the base $\mathcal{E}_{EG-SW}$ can be written as

$$\begin{aligned}\mathcal{E}_{EG-SW} &= \mathcal{E}_{E-G} + \mathcal{E}_{S-W} \\ &= (\mathbf{1}, \vec{e}_1, \vec{e}_2, \vec{e}_3, \vec{e}_4, \vec{e}_5, \vec{e}_6, \vec{e}_7, \vec{e}_8, \vec{e}_9, \vec{e}_{10}, \vec{e}_{11}, \vec{e}_{12}, \vec{e}_{13}, \vec{e}_{14}, \vec{e}_{15})\end{aligned} \tag{33}$$

The displacement ($r_0, r_1, r_2, r_3, r_4, r_5, r_6, r_7, r_8, r_9, r_{10}, r_{11}, r_{12}, r_{13}, r_{14}, r_{15}$) in the sedenion space EG-SW is

$$\begin{aligned}\mathcal{R}_{EG-SW} =&\ \mathcal{R}_{E-G} + \mathcal{R}_{S-W} \\ =&\ r_0 + \vec{e}_1 r_1 + \vec{e}_2 r_2 + \vec{e}_3 r_3 + \vec{e}_4 r_4 + \vec{e}_5 r_5 + \vec{e}_6 r_6 + \vec{e}_7 r_7 + \vec{e}_8 r_8 \\ &+ \vec{e}_9 r_9 + \vec{e}_{10} r_{10} + \vec{e}_{11} r_{11} + \vec{e}_{12} r_{12} + \vec{e}_{13} r_{13} + \vec{e}_{14} r_{14} + \vec{e}_{15} r_{15}\end{aligned} \tag{34}$$

The sedenion differential operator $\diamondsuit_{EGSW}$ and its conjugate operator are defined as,

$$\diamondsuit_{EGSW} = \diamondsuit_{EGSW1} + \diamondsuit_{EGSW2}\ ,\qquad \diamondsuit^*_{EGSW} = \diamondsuit^*_{EGSW1} + \diamondsuit^*_{EGSW2} \tag{35}$$

In the sedenion field EG-SW, the definition of the field source is very different to that of the octonionic field, and thus the succeeding equations and operators should be revised. By analogy with the case of other fields, the sedenion differential operator $\diamondsuit$ needs to be generalized to the new operator ($\mathcal{B}_{E-G} / k^B_{E-G} + \mathcal{B}_{S-W} / k^B_{S-W} + \diamondsuit$). This is because the sedenion field EG-SW includes electromagnetic-gravitational field and strong-weak field. Therefore the physical characteristics of the sedenion field EG-SW will be much more complicated than that of the octonionic fields in many aspects.

It can be predicted that the field strength of off-diagonal subfields may be symmetry along the diagonal subfield in Table 7. For example, the field strength of the weak-electromagnetic subfield may be equal to that of the electromagnetic-weak subfield. And it is easy to find that weak interaction is interconnected with electromagnetic interaction. The physical features of each subfield in the sedenion field EG-SW meet the requirement of equations set in Table 8.

Table 7. Subfields of sedenion field EG-SW

|  | Strong Interaction | Weak Interaction | Electromagnetic Interaction | Gravitational Interaction |
|---|---|---|---|---|
| strong quaternionic space S | strong-strong subfield S strong-charge $\gamma^S_{ss}$ | weak-strong subfield S weak-charge $\gamma^S_{ws}$ | electromagnetic-strong subfield S electromagnetic-charge, $\gamma^S_{es}$ | gravitational-strong subfield S gravitational-charge, $\gamma^S_{gs}$ |
| weak quaternionic space W | strong-weak subfield W strong-charge $\gamma^S_{sw}$ | weak-weak subfield W weak-charge $\gamma^S_{ww}$ | electromagnetic-weak subfield W electromagnetic-charge, $\gamma^S_{ew}$ | gravitational-weak subfield, W gravitational-charge, $\gamma^S_{gw}$ |
| electromagnetic quaternionic space E | strong-electromagnetic subfield E strong-charge, $\gamma^S_{se}$ | weak-electromagnetic subfield E weak-charge, $\gamma^S_{we}$ | electromagnetic-electromagnetic subfield, E electromagnetic-charge, $\gamma^S_{ee}$ | gravitational-electromagnetic subfield, E gravitational-charge, $\gamma^S_{ge}$ |
| gravitational quaternionic space G | strong-gravitational subfield G strong-charge, $\gamma^S_{sg}$ | weak-gravitational subfield G weak-charge, $\gamma^S_{wg}$ | electromagnetic-gravitational subfield, G electromagnetic-charge, $\gamma^S_{eg}$ | gravitational-gravitational subfield, G gravitational-charge, $\gamma^S_{gg}$ |

In the sedenion field EG-SW, the field potential ($a_0$, $a_1$, $a_2$, $a_3$, $a_4$, $a_5$, $a_6$, $a_7$, $a_8$, $a_9$, $a_{10}$, $a_{11}$, $a_{12}$, $a_{13}$, $a_{14}$, $a_{15}$) is defined as

$$\begin{aligned}\mathcal{A} &= \diamondsuit^* \circ \mathcal{X} \\ &= a_0 + a_1 \vec{e}_1 + a_2 \vec{e}_2 + a_3 \vec{e}_3 + a_4 \vec{e}_4 + a_5 \vec{e}_5 \\ &\quad + a_6 \vec{e}_6 + a_7 \vec{e}_7 + a_8 \vec{e}_8 + a_9 \vec{e}_9 + a_{10} \vec{e}_{10} \\ &\quad + a_{11} \vec{e}_{11} + a_{12} \vec{e}_{12} + a_{13} \vec{e}_{13} + a_{14} \vec{e}_{14} + a_{15} \vec{e}_{15}\end{aligned} \quad (36)$$

where, $k_{rx} \mathcal{X} = k^s_{rx} \mathcal{X}_s + k^w_{rx} \mathcal{X}_w + k^e_{rx} \mathcal{X}_e + k^g_{rx} \mathcal{X}_g$. $\mathcal{X}_s$ is physical quantity in S space, $\mathcal{X}_w$ is physical quantity in W space, $\mathcal{X}_e$ is physical quantity in E space, and $\mathcal{X}_g$ is physical quantity in G space. $k_{rx}$, $k^s_{rx}$, $k^w_{rx}$, $k^e_{rx}$ and $k^g_{rx}$ are coefficients.

The field strength $\mathcal{B}$ of the sedenion field EG-SW can be defined as

$$\mathcal{B} = \diamondsuit \circ \mathcal{A} \qquad (37)$$

Then the field source and the force of the sedenion field can be defined respectively as

$$\mu \mathcal{S} = (\mathcal{B}/K + \diamondsuit)^* \circ \mathcal{B} = (\mathcal{B}_{E\text{-}G}/k^B_{E\text{-}G} + \mathcal{B}_{S\text{-}W}/k^B_{S\text{-}W} + \diamondsuit)^* \circ \mathcal{B} \qquad (38)$$

$$\mathcal{Z} = K(\mathcal{B}_{E\text{-}G}/k^B_{E\text{-}G} + \mathcal{B}_{S\text{-}W}/k^B_{S\text{-}W} + \diamondsuit) \circ \mathcal{S} \qquad (39)$$

where, $k^B_{S\text{-}W}$ and $k^B_{E\text{-}G}$ are coefficients. $K = K^B_{EGSW}$ is the coefficient. $\mathcal{B}_{S\text{-}W}$ is the field strength in strong-weak field, and $\mathcal{B}_{E\text{-}G}$ is the field strength in electromagnetic-gravitational field. The mark (*) denotes sedenion conjugate. The coefficient μ is interaction intensity of the sedenion field EG-SW.

The angular momentum of the sedenion field can be defined as

$$\mathcal{M} = \mathcal{S} \circ (\mathcal{R} + k_{rx}\mathcal{X}) \qquad (40)$$

and the energy and power in the sedenion field can be defined respectively as

$$\mathcal{W} = K(\mathcal{B}_{E\text{-}G}/k^B_{E\text{-}G} + \mathcal{B}_{S\text{-}W}/k^B_{S\text{-}W} + \diamondsuit)^* \circ \mathcal{M} \qquad (41)$$

$$\mathcal{N} = K(\mathcal{B}_{E\text{-}G}/k^B_{E\text{-}G} + \mathcal{B}_{S\text{-}W}/k^B_{S\text{-}W} + \diamondsuit) \circ \mathcal{W} \qquad (42)$$

Table 8. Equations set of sedenion field EG-SW

| Spacetime | Sedenion space EG-SW |
|---|---|
| $\mathcal{X}$ physical quantity | $k_{rx}\mathcal{X} = k^s_{rx}\mathcal{X}_s + k^w_{rx}\mathcal{X}_w + k^e_{rx}\mathcal{X}_e + k^g_{rx}\mathcal{X}_g$ |
| Field potential | $\mathcal{A} = \diamondsuit^* \circ \mathcal{X}$ |
| Field strength | $\mathcal{B} = \diamondsuit \circ \mathcal{A}$ |
| Field source | $K\mu\mathcal{S} = K(\mathcal{B}_{E\text{-}G}/k^B_{E\text{-}G} + \mathcal{B}_{S\text{-}W}/k^B_{S\text{-}W} + \diamondsuit)^* \circ \mathcal{B}$ |
| Force | $\mathcal{Z} = K(\mathcal{B}_{E\text{-}G}/k^B_{E\text{-}G} + \mathcal{B}_{S\text{-}W}/k^B_{S\text{-}W} + \diamondsuit) \circ \mathcal{S}$ |
| Angular momentum | $\mathcal{M} = \mathcal{S} \circ (\mathcal{R} + k_{rx}\mathcal{X})$ |
| Energy | $\mathcal{W} = K(\mathcal{B}_{E\text{-}G}/k^B_{E\text{-}G} + \mathcal{B}_{S\text{-}W}/k^B_{S\text{-}W} + \diamondsuit)^* \circ \mathcal{M}$ |
| Power | $\mathcal{N} = K(\mathcal{B}_{E\text{-}G}/k^B_{E\text{-}G} + \mathcal{B}_{S\text{-}W}/k^B_{S\text{-}W} + \diamondsuit) \circ \mathcal{W}$ |

In the sedenion space EG-SW, the wave functions of quantum mechanics are the sedenion equations set. The Dirac and Klein-Gordon equations of the quantum mechanics are actually the wave equations set which are associated with the particle's angular momentum.

In sedenion field EG-SW, the Dirac equation and Klein-Gordon equation can be attained respectively from the energy equation (41) and power equation (42) after substituting operator $K(\mathcal{B}_{E\text{-}G}/k^B_{E\text{-}G} + \mathcal{B}_{S\text{-}W}/k^B_{S\text{-}W} + \diamondsuit)$ for $(\mathcal{W}_{E\text{-}G}/k^B_{E\text{-}G} b^B_{E\text{-}G} + \mathcal{W}_{S\text{-}W}/k^B_{S\text{-}W} b^B_{S\text{-}W} + \diamondsuit)$. The coefficients $b^B_{E\text{-}G}$ and $b^B_{S\text{-}W}$ are the Plank-like constant.

The $\mathcal{U}$ equation of the quantum mechanics can be defined as

$$\mathcal{U} = (\mathcal{W}_{E\text{-}G}/k^B_{E\text{-}G} b^B_{E\text{-}G} + \mathcal{W}_{S\text{-}W}/k^B_{S\text{-}W} b^B_{S\text{-}W} + \diamondsuit)^* \circ \mathcal{M} \qquad (43)$$

The $\mathcal{L}$ equation of the quantum mechanics can be defined as

$$\mathcal{L} = (\mathcal{W}_{E\text{-}G}/k^B_{E\text{-}G} b^B_{E\text{-}G} + \mathcal{W}_{S\text{-}W}/k^B_{S\text{-}W} b^B_{S\text{-}W} + \diamondsuit) \circ \mathcal{U} \qquad (44)$$

The Dirac-like equations can be obtained from the field source equation (38) and force equation (39) respectively.

The $\mathcal{T}$ equation of the quantum mechanics can be defined as

$$\mathcal{T} = (\mathcal{W}_{E\text{-}G}/k^B_{E\text{-}G} b^B_{E\text{-}G} + \mathcal{W}_{S\text{-}W}/k^B_{S\text{-}W} b^B_{S\text{-}W} + \diamondsuit)^* \circ \mathcal{B} \qquad (45)$$

The $\mathcal{O}$ equation of the quantum mechanics can be defined as

$$O = (\mathcal{W}_{E\text{-}G} / k^B{}_{E\text{-}G} b^B{}_{E\text{-}G} + \mathcal{W}_{S\text{-}W} / k^B{}_{S\text{-}W} b^B{}_{S\text{-}W} + \diamondsuit) \circ \mathcal{T} \qquad (46)$$

Table 9.   Quantum equations set of sedenion field EG-SW

| | |
|---|---|
| Energy quantum | $\mathcal{U} = (\mathcal{W}_{E\text{-}G} / k^B{}_{E\text{-}G} b^B{}_{E\text{-}G} + \mathcal{W}_{S\text{-}W} / k^B{}_{S\text{-}W} b^B{}_{S\text{-}W} + \diamondsuit)^* \circ \mathcal{M}$ |
| Power quantum | $\mathcal{L} = (\mathcal{W}_{E\text{-}G} / k^B{}_{E\text{-}G} b^B{}_{E\text{-}G} + \mathcal{W}_{S\text{-}W} / k^B{}_{S\text{-}W} b^B{}_{S\text{-}W} + \diamondsuit) \circ \mathcal{U}$ |
| Field source quantum | $\mathcal{T} = (\mathcal{W}_{E\text{-}G} / k^B{}_{E\text{-}G} b^B{}_{E\text{-}G} + \mathcal{W}_{S\text{-}W} / k^B{}_{S\text{-}W} b^B{}_{S\text{-}W} + \diamondsuit)^* \circ \mathcal{B}$ |
| Force quantum | $O = (\mathcal{W}_{E\text{-}G} / k^B{}_{E\text{-}G} b^B{}_{E\text{-}G} + \mathcal{W}_{S\text{-}W} / k^B{}_{S\text{-}W} b^B{}_{S\text{-}W} + \diamondsuit) \circ \mathcal{T}$ |

In the sedenion field EG-SW, the intermediate and field source particles can be gained. We can find the intermediate particles and other kinds of new and unknown particles which may be existed in the nature.

## 5. Compounding field in sedenion space HS-SW

It is believed that the hyper-strong field and the strong-weak field are unified, equal and interconnected. By means of the conception of the space expansion etc., two types of the octonionic spaces (Octonionic space H-S and Octonionic space S-W) can be combined into a sedenion space (Sedenion space HS-SW, for short). In the sedenion space HS-SW, some properties of hyper-strong field and strong-weak field can be described uniformly.

In the sedenion space HS-SW, there exist one kind of field (sedenion field HS-SW, for short) can be obtained related to the operator ($\mathcal{X}/k + \diamondsuit$).

In the octonionic space H-S, the base $\mathcal{E}_{H\text{-}S}$ can be written as
$$\mathcal{E}_{H\text{-}S} = (\mathbf{1}, \vec{e}_1, \vec{e}_2, \vec{e}_3, \vec{e}_4, \vec{e}_5, \vec{e}_6, \vec{e}_7) \qquad (47)$$

The displacement ($r_0, r_1, r_2, r_3, r_4, r_5, r_6, r_7$) in the octonionic space H-S is
$$\mathcal{R}_{H\text{-}S} = r_0 + \vec{e}_1 r_1 + \vec{e}_2 r_2 + \vec{e}_3 r_3 + \vec{e}_4 r_4 + \vec{e}_5 r_5 + \vec{e}_6 r_6 + \vec{e}_7 r_7 \qquad (48)$$

where, $r_0 = c_{H\text{-}S} t_{H\text{-}S}$, $r_4 = c_{H\text{-}S} T_{H\text{-}S}$. $c_{H\text{-}S}$ is the speed of intermediate particle in the hyper-strong field, $t_{H\text{-}S}$ and $T_{H\text{-}S}$ denote the time.

The octonionic differential operator $\diamondsuit_{HSSW1}$ and its conjugate operator are defined as
$$\diamondsuit_{HSSW1} = \partial_0 + \vec{e}_1 \partial_1 + \vec{e}_2 \partial_2 + \vec{e}_3 \partial_3 + \vec{e}_4 \partial_4 + \vec{e}_5 \partial_5 + \vec{e}_6 \partial_6 + \vec{e}_7 \partial_7 \qquad (49)$$
$$\diamondsuit^*_{HSSW1} = \partial_0 - \vec{e}_1 \partial_1 - \vec{e}_2 \partial_2 - \vec{e}_3 \partial_3 - \vec{e}_4 \partial_4 - \vec{e}_5 \partial_5 - \vec{e}_6 \partial_6 - \vec{e}_7 \partial_7 \qquad (50)$$

where, $\partial_j = \partial / \partial r_j$, j = 0, 1, 2, 3, 4, 5, 6, 7.

In the octonionic space S-W, the base $\mathcal{E}_{S\text{-}W}$ can be written as
$$\mathcal{E}_{S\text{-}W} = (\vec{e}_8, \vec{e}_9, \vec{e}_{10}, \vec{e}_{11}, \vec{e}_{12}, \vec{e}_{13}, \vec{e}_{14}, \vec{e}_{15}) \qquad (51)$$

The displacement ($r_8, r_9, r_{10}, r_{11}, r_{12}, r_{13}, r_{14}, r_{15}$) in the octonionic space S-W is
$$\mathcal{R}_{S\text{-}W} = \vec{e}_8 r_8 + \vec{e}_9 r_9 + \vec{e}_{10} r_{10} + \vec{e}_{11} r_{11} + \vec{e}_{12} r_{12} + \vec{e}_{13} r_{13} + \vec{e}_{14} r_{14} + \vec{e}_{15} r_{15} \qquad (52)$$

where, $r_8 = c_{S\text{-}W} t_{S\text{-}W}$, $r_{12} = c_{S\text{-}W} T_{S\text{-}W}$. $c_{S\text{-}W}$ is the speed of intermediate particle in the strong-weak field, $t_{S\text{-}W}$ and $T_{S\text{-}W}$ denote the time.

The octonionic differential operator $\diamondsuit_{HSSW2}$ and its conjugate operator are defined as,
$$\diamondsuit_{HSSW2} = \vec{e}_8 \partial_8 + \vec{e}_9 \partial_9 + \vec{e}_{10} \partial_{10} + \vec{e}_{11} \partial_{11}$$
$$+ \vec{e}_{12} \partial_{12} + \vec{e}_{13} \partial_{13} + \vec{e}_{14} \partial_{14} + \vec{e}_{15} \partial_{15} \qquad (53)$$
$$\diamondsuit^*_{HSSW2} = -\vec{e}_8 \partial_8 - \vec{e}_9 \partial_9 - \vec{e}_{10} \partial_{10} - \vec{e}_{11} \partial_{11}$$
$$- \vec{e}_{12} \partial_{12} - \vec{e}_{13} \partial_{13} - \vec{e}_{14} \partial_{14} - \vec{e}_{15} \partial_{15} \qquad (54)$$

where, $\partial_j = \partial / \partial r_j$ , j = 8, 9, 10, 11, 12, 13, 14, 15 .

In the sedenion space HS-SW, the base $\mathcal{E}_{\text{HS-SW}}$ can be written as

$$\mathcal{E}_{\text{HS-SW}} = \mathcal{E}_{\text{H-S}} + \mathcal{E}_{\text{S-W}}$$
$$= (1, \vec{e}_1, \vec{e}_2, \vec{e}_3, \vec{e}_4, \vec{e}_5, \vec{e}_6, \vec{e}_7, \vec{e}_8, \vec{e}_9, \vec{e}_{10}, \vec{e}_{11}, \vec{e}_{12}, \vec{e}_{13}, \vec{e}_{14}, \vec{e}_{15}) \quad (55)$$

The displacement ( $r_0$ , $r_1$ , $r_2$ , $r_3$ , $r_4$ , $r_5$ , $r_6$ , $r_7$ , $r_8$ , $r_9$ , $r_{10}$ , $r_{11}$ , $r_{12}$ , $r_{13}$ , $r_{14}$ , $r_{15}$ ) in the sedenion space HS-SW is

$$\mathcal{R}_{\text{HS-SW}} = \mathcal{R}_{\text{H-S}} + \mathcal{R}_{\text{S-W}}$$
$$= r_0 + \vec{e}_1 r_1 + \vec{e}_2 r_2 + \vec{e}_3 r_3 + \vec{e}_4 r_4 + \vec{e}_5 r_5 + \vec{e}_6 r_6 + \vec{e}_7 r_7 + \vec{e}_8 r_8$$
$$+ \vec{e}_9 r_9 + \vec{e}_{10} r_{10} + \vec{e}_{11} r_{11} + \vec{e}_{12} r_{12} + \vec{e}_{13} r_{13} + \vec{e}_{14} r_{14} + \vec{e}_{15} r_{15} \quad (56)$$

The sedenion differential operator $\diamondsuit_{\text{HSSW}}$ and its conjugate operator are defined as,

$$\diamondsuit_{\text{HSSW}} = \diamondsuit_{\text{HSSW1}} + \diamondsuit_{\text{HSSW2}} \; , \quad \diamondsuit^*_{\text{HSSW}} = \diamondsuit^*_{\text{HSSW1}} + \diamondsuit^*_{\text{HSSW2}} \quad (57)$$

Table 10.   Subfields of the sedenion field HS-SW

|  | A Interaction | B Interaction | Strong Interaction | Weak Interaction |
|---|---|---|---|---|
| A quaternionic space A | A-A subfield<br>A A-charge<br>$\gamma^A_{aa}$ | B-A subfield<br>A B-charge<br>$\gamma^A_{ba}$ | strong-A subfield<br>A strong-charge<br>$\gamma^A_{sa}$ | weak-A subfield<br>A weak-charge<br>$\gamma^A_{wa}$ |
| B quaternionic space B | A-B subfield<br>B A-charge<br>$\gamma^A_{ab}$ | B-B subfield<br>B B-charge<br>$\gamma^A_{bb}$ | strong-B subfield<br>B strong-charge<br>$\gamma^A_{sb}$ | weak-B subfield<br>B weak-charge<br>$\gamma^A_{wb}$ |
| strong quaternionic space S | A-strong subfield<br>S A-charge<br>$\gamma^A_{as}$ | B-strong subfield<br>S B-charge<br>$\gamma^A_{bs}$ | strong-strong subfield<br>S strong-charge<br>$\gamma^A_{ss}$ | weak-strong subfield,<br>S weak-charge<br>$\gamma^A_{ws}$ |
| weak quaternionic space W | A-weak subfield<br>W A-charge<br>$\gamma^A_{aw}$ | B-weak subfield<br>W B-charge<br>$\gamma^A_{bw}$ | strong-weak subfield<br>W strong-charge<br>$\gamma^A_{sw}$ | weak-weak subfield<br>W weak-charge<br>$\gamma^A_{ww}$ |

In the sedenion field HS-SW, the definition of the field potential is very different to that of the octonionic field, and thus the succeeding equations and operators should be revised. By analogy with the case of other fields, the sedenion differential operator $\diamondsuit$ needs to be generalized to the new operator ( $\mathcal{X}_{\text{H-S}} / k^X_{\text{H-S}} + \mathcal{X}_{\text{S-W}} / k^X_{\text{S-W}} + \diamondsuit$ ). This is because the sedenion field HS-SW includes the hyper-strong field and strong-weak field. Therefore the physical characteristics of the sedenion field HS-SW will be much more complicated than that of the octonionic fields in many aspects.

It can be predicted that the field strength of off-diagonal subfields may be symmetry along the diagonal subfield in the Table 10. The hyper-strong consists of A and B interactions. For example, the field strength of the B-strong subfield may be equal to that of the strong-B subfield. And it is easy to find that B interaction is interconnected with strong interaction. The physical features of each subfield in the sedenion field HS-SW meet the requirement of the equations set in the Table 11.

In the sedenion field HS-SW, the field potential ($a_0$, $a_1$, $a_2$, $a_3$, $a_4$, $a_5$, $a_6$, $a_7$, $a_8$, $a_9$, $a_{10}$, $a_{11}$, $a_{12}$, $a_{13}$, $a_{14}$, $a_{15}$) is defined as

$$\mathcal{A} = (\mathcal{X}/K + \diamondsuit)^* \circ \mathcal{X} = (\mathcal{X}_{H-S}/k^X_{H-S} + \mathcal{X}_{S-W}/k^X_{S-W} + \diamondsuit)^* \circ \mathcal{X}$$
$$= a_0 + a_1 \vec{e}_1 + a_2 \vec{e}_2 + a_3 \vec{e}_3 + a_4 \vec{e}_4 + a_5 \vec{e}_5$$
$$+ a_6 \vec{e}_6 + a_7 \vec{e}_7 + a_8 \vec{e}_8 + a_9 \vec{e}_9 + a_{10} \vec{e}_{10}$$
$$+ a_{11} \vec{e}_{11} + a_{12} \vec{e}_{12} + a_{13} \vec{e}_{13} + a_{14} \vec{e}_{14} + a_{15} \vec{e}_{15} \qquad (58)$$

where, $k_{rx}\mathcal{X} = k^{S-W}_{rx}\mathcal{X}_{S-W} + k^{H-S}_{rx}\mathcal{X}_{H-S}$. $k^{S-W}_{rx}\mathcal{X}_{S-W} = k^s_{rx}\mathcal{X}_s + k^w_{rx}\mathcal{X}_w$. And $K$, $k^{S-W}_{rx}$, $k^{H-S}_{rx}$, $k^X_{S-W}$ and $k^X_{H-S}$ are coefficients. And $\mathcal{X}_{H-S}$ is physical quantity in hyper-strong field.

The field strength $\mathcal{B}$ of the sedenion field HS-SW can be defined as

$$\mathcal{B} = (\mathcal{X}_{H-S}/k^X_{H-S} + \mathcal{X}_{S-W}/k^X_{S-W} + \diamondsuit) \circ \mathcal{A} \qquad (59)$$

Then the field source and the force of the sedenion field can be defined respectively as

$$K\mu S = K(\mathcal{X}_{H-S}/k^X_{H-S} + \mathcal{X}_{S-W}/k^X_{S-W} + \diamondsuit)^* \circ \mathcal{B} \qquad (60)$$
$$\mathcal{Z} = K(\mathcal{X}_{H-S}/k^X_{H-S} + \mathcal{X}_{S-W}/k^X_{S-W} + \diamondsuit) \circ S \qquad (61)$$

where, the mark (*) denotes sedenion conjugate. The coefficient $\mu$ is interaction intensity of the sedenion field HS-SW. $K = K^X_{HSSW}$ is the coefficient.

The angular momentum of the sedenion field can be defined as

$$\mathcal{M} = S \circ (\mathcal{R} + k_{rx}\mathcal{X}) \qquad (62)$$

and the energy and power in the sedenion field can be defined respectively as

$$\mathcal{W} = K(\mathcal{X}_{H-S}/k^X_{H-S} + \mathcal{X}_{S-W}/k^X_{S-W} + \diamondsuit)^* \circ \mathcal{M} \qquad (63)$$
$$\mathcal{N} = K(\mathcal{X}_{H-S}/k^X_{H-S} + \mathcal{X}_{S-W}/k^X_{S-W} + \diamondsuit) \circ \mathcal{W} \qquad (64)$$

Table 11.  Equations set of sedenion field HS-SW

| Spacetime | Sedenion space HS-SW |
| --- | --- |
| $\mathcal{X}$ physical quantity | $k_{rx}\mathcal{X} = k^{S-W}_{rx}\mathcal{X}_{S-W} + k^{H-S}_{rx}\mathcal{X}_{H-S}$ |
| Field potential | $\mathcal{A} = (\mathcal{X}_{H-S}/k^X_{H-S} + \mathcal{X}_{S-W}/k^X_{S-W} + \diamondsuit)^* \circ \mathcal{X}$ |
| Field strength | $\mathcal{B} = (\mathcal{X}_{H-S}/k^X_{H-S} + \mathcal{X}_{S-W}/k^X_{S-W} + \diamondsuit) \circ \mathcal{A}$ |
| Field source | $K\mu S = K(\mathcal{X}_{H-S}/k^X_{H-S} + \mathcal{X}_{S-W}/k^X_{S-W} + \diamondsuit)^* \circ \mathcal{B}$ |
| Force | $\mathcal{Z} = K(\mathcal{X}_{H-S}/k^X_{H-S} + \mathcal{X}_{S-W}/k^X_{S-W} + \diamondsuit) \circ S$ |
| Angular momentum | $\mathcal{M} = S \circ (\mathcal{R} + k_{rx}\mathcal{X})$ |
| Energy | $\mathcal{W} = K(\mathcal{X}_{H-S}/k^X_{H-S} + \mathcal{X}_{S-W}/k^X_{S-W} + \diamondsuit)^* \circ \mathcal{M}$ |
| Power | $\mathcal{N} = K(\mathcal{X}_{H-S}/k^X_{H-S} + \mathcal{X}_{S-W}/k^X_{S-W} + \diamondsuit) \circ \mathcal{W}$ |

In sedenion field HS-SW, the Dirac equation and Klein-Gordon equation can be attained respectively from the energy equation (63) and power equation (64) after substituting operator $K(\mathcal{X}_{H-S}/k^X_{H-S} + \mathcal{X}_{S-W}/k^X_{S-W} + \diamondsuit)$ for $(\mathcal{W}_{H-S}/k^X_{H-S}b^X_{H-S} + \mathcal{W}_{S-W}/k^X_{S-W}b^X_{S-W} + \diamondsuit)$. The coefficients $b^X_{H-S}$ and $b^X_{S-W}$ are the Plank-like constant.

The $\mathcal{U}$ equation of the quantum mechanics can be defined as

$$\mathcal{U} = (\mathcal{W}_{H-S} / k^X_{H-S} b^X_{H-S} + \mathcal{W}_{S-W} / k^X_{S-W} b^X_{S-W} + \diamond)^* \circ \mathcal{M} \quad (65)$$

The $\mathcal{L}$ equation of the quantum mechanics can be defined as

$$\mathcal{L} = (\mathcal{W}_{H-S} / k^X_{H-S} b^X_{H-S} + \mathcal{W}_{S-W} / k^X_{S-W} b^X_{S-W} + \diamond) \circ \mathcal{U} \quad (66)$$

The four sorts of Dirac-like equations can be obtained from the Eqs.(58), (59), (60) and (61) respectively.

The $\mathcal{D}$ equation of the quantum mechanics can be defined as

$$\mathcal{D} = (\mathcal{W}_{H-S} / k^X_{H-S} b^X_{H-S} + \mathcal{W}_{S-W} / k^X_{S-W} b^X_{S-W} + \diamond)^* \circ \mathcal{X} \quad (67)$$

The $\mathcal{G}$ equation of the quantum mechanics can be defined as

$$\mathcal{G} = (\mathcal{W}_{H-S} / k^X_{H-S} b^X_{H-S} + \mathcal{W}_{S-W} / k^X_{S-W} b^X_{S-W} + \diamond) \circ \mathcal{D} \quad (68)$$

The $\mathcal{T}$ equation of the quantum mechanics can be defined as

$$\mathcal{T} = (\mathcal{W}_{H-S} / k^X_{H-S} b^X_{H-S} + \mathcal{W}_{S-W} / k^X_{S-W} b^X_{S-W} + \diamond)^* \circ \mathcal{G} \quad (69)$$

The $\mathcal{O}$ equation of the quantum mechanics can be defined as

$$\mathcal{O} = (\mathcal{W}_{H-S} / k^X_{H-S} b^X_{H-S} + \mathcal{W}_{S-W} / k^X_{S-W} b^X_{S-W} + \diamond) \circ \mathcal{T} \quad (70)$$

Table 12.   Quantum equations set of sedenion field HS-SW

| | |
|---|---|
| Energy quantum | $\mathcal{U} = (\mathcal{W}_{H-S} / k^X_{H-S} b^X_{H-S} + \mathcal{W}_{S-W} / k^X_{S-W} b^X_{S-W} + \diamond)^* \circ \mathcal{M}$ |
| Power quantum | $\mathcal{L} = (\mathcal{W}_{H-S} / k^X_{H-S} b^X_{H-S} + \mathcal{W}_{S-W} / k^X_{S-W} b^X_{S-W} + \diamond) \circ \mathcal{U}$ |
| Field potential quantum | $\mathcal{D} = (\mathcal{W}_{H-S} / k^X_{H-S} b^X_{H-S} + \mathcal{W}_{S-W} / k^X_{S-W} b^X_{S-W} + \diamond)^* \circ \mathcal{X}$ |
| Field strength quantum | $\mathcal{G} = (\mathcal{W}_{H-S} / k^X_{H-S} b^X_{H-S} + \mathcal{W}_{S-W} / k^X_{S-W} b^X_{S-W} + \diamond) \circ \mathcal{D}$ |
| Field source quantum | $\mathcal{T} = (\mathcal{W}_{H-S} / k^X_{H-S} b^X_{H-S} + \mathcal{W}_{S-W} / k^X_{S-W} b^X_{S-W} + \diamond)^* \circ \mathcal{G}$ |
| Force quantum | $\mathcal{O} = (\mathcal{W}_{H-S} / k^X_{H-S} b^X_{H-S} + \mathcal{W}_{S-W} / k^X_{S-W} b^X_{S-W} + \diamond) \circ \mathcal{T}$ |

In the sedenion field HS-SW, the intermediate and field source particles can be drawn. We can find the intermediate particles and other kinds of new and unknown particles which may be existed in the nature.

## 6. Compounding field in sedenion space SW-HS

It is believed that the strong-weak field and the hyper-strong field are unified, equal and interconnected. By means of the conception of the space expansion etc., two types of the octonionic spaces (Octonionic space S-W and Octonionic space H-S) can be combined into a sedenion space (Sedenion space SW-HS, for short), which is different to the sedenion space HS-SW. In the sedenion space SW-HS, some properties of the hyper-strong field and the strong-weak field can be described uniformly.

In sedenion space SW-HS, there exist one kind of field (sedenion field SW-HS, for short) which is different to sedenion field HS-SW, can be obtained related to operator ($\mathcal{A}/k + \diamond$).

In the octonionic space S-W, the base $\mathcal{E}_{S-W}$ can be written as

$$\mathcal{E}_{S-W} = (\mathbf{1},\ \vec{e}_1,\ \vec{e}_2,\ \vec{e}_3,\ \vec{e}_4,\ \vec{e}_5,\ \vec{e}_6,\ \vec{e}_7) \quad (71)$$

The displacement ($r_0$, $r_1$, $r_2$, $r_3$, $r_4$, $r_5$, $r_6$, $r_7$) in the octonionic space S-W is

$$\mathcal{R}_{S-W} = r_0 + \vec{e}_1 r_1 + \vec{e}_2 r_2 + \vec{e}_3 r_3 + \vec{e}_4 r_4 + \vec{e}_5 r_5 + \vec{e}_6 r_6 + \vec{e}_7 r_7 \quad (72)$$

where, $r_0 = c_{S-W} t_{S-W}$, $r_4 = c_{S-W} T_{S-W}$. $c_{S-W}$ is the speed of intermediate particle in the strong-weak field, $t_{S-W}$ and $T_{S-W}$ denote the time.

The octonionic differential operator $\diamondsuit_{SWHS1}$ and its conjugate operator are defined as

$$\diamondsuit_{SWHS1} = \partial_0 + \vec{e}_1 \partial_1 + \vec{e}_2 \partial_2 + \vec{e}_3 \partial_3 + \vec{e}_4 \partial_4 + \vec{e}_5 \partial_5 + \vec{e}_6 \partial_6 + \vec{e}_7 \partial_7 \tag{73}$$

$$\diamondsuit^*_{SWHS1} = \partial_0 - \vec{e}_1 \partial_1 - \vec{e}_2 \partial_2 - \vec{e}_3 \partial_3 - \vec{e}_4 \partial_4 - \vec{e}_5 \partial_5 - \vec{e}_6 \partial_6 - \vec{e}_7 \partial_7 \tag{74}$$

where, $\partial_j = \partial/\partial r_j$, $j = 0, 1, 2, 3, 4, 5, 6, 7$.

In the octonionic space H-S, the base $\mathcal{E}_{H-S}$ can be written as

$$\mathcal{E}_{H-S} = (\vec{e}_8, \vec{e}_9, \vec{e}_{10}, \vec{e}_{11}, \vec{e}_{12}, \vec{e}_{13}, \vec{e}_{14}, \vec{e}_{15}) \tag{75}$$

The displacement ($r_8, r_9, r_{10}, r_{11}, r_{12}, r_{13}, r_{14}, r_{15}$) in the octonionic space H-S is

$$\mathcal{R}_{H-S} = \vec{e}_8 r_8 + \vec{e}_9 r_9 + \vec{e}_{10} r_{10} + \vec{e}_{11} r_{11} + \vec{e}_{12} r_{12} + \vec{e}_{13} r_{13} + \vec{e}_{14} r_{14} + \vec{e}_{15} r_{15} \tag{76}$$

where, $r_8 = c_{H-S} t_{H-S}$, $r_{12} = c_{H-S} T_{H-S}$. $c_{H-S}$ is the speed of intermediate particle in the hyper-strong field, $t_{H-S}$ and $T_{H-S}$ denote the time.

The octonionic differential operator $\diamondsuit_{SWHS2}$ and its conjugate operator are defined as,

$$\diamondsuit_{SWHS2} = \vec{e}_8 \partial_8 + \vec{e}_9 \partial_9 + \vec{e}_{10} \partial_{10} + \vec{e}_{11} \partial_{11}$$
$$+ \vec{e}_{12} \partial_{12} + \vec{e}_{13} \partial_{13} + \vec{e}_{14} \partial_{14} + \vec{e}_{15} \partial_{15} \tag{77}$$

$$\diamondsuit^*_{SWHS2} = -\vec{e}_8 \partial_8 - \vec{e}_9 \partial_9 - \vec{e}_{10} \partial_{10} - \vec{e}_{11} \partial_{11}$$
$$- \vec{e}_{12} \partial_{12} - \vec{e}_{13} \partial_{13} - \vec{e}_{14} \partial_{14} - \vec{e}_{15} \partial_{15} \tag{78}$$

where, $\partial_j = \partial/\partial r_j$, $j = 8, 9, 10, 11, 12, 13, 14, 15$.

In the sedenion space SW-HS, the base $\mathcal{E}_{SW-HS}$ can be written as

$$\mathcal{E}_{SW-HS} = \mathcal{E}_{S-W} + \mathcal{E}_{H-S}$$
$$= (1, \vec{e}_1, \vec{e}_2, \vec{e}_3, \vec{e}_4, \vec{e}_5, \vec{e}_6, \vec{e}_7, \vec{e}_8, \vec{e}_9, \vec{e}_{10}, \vec{e}_{11}, \vec{e}_{12}, \vec{e}_{13}, \vec{e}_{14}, \vec{e}_{15}) \tag{79}$$

The displacement ($r_0, r_1, r_2, r_3, r_4, r_5, r_6, r_7, r_8, r_9, r_{10}, r_{11}, r_{12}, r_{13}, r_{14}, r_{15}$) in the sedenion space SW-HS is

$$\mathcal{R}_{SW-HS} = \mathcal{R}_{S-W} + \mathcal{R}_{H-S}$$
$$= r_0 + \vec{e}_1 r_1 + \vec{e}_2 r_2 + \vec{e}_3 r_3 + \vec{e}_4 r_4 + \vec{e}_5 r_5 + \vec{e}_6 r_6 + \vec{e}_7 r_7 + \vec{e}_8 r_8$$
$$+ \vec{e}_9 r_9 + \vec{e}_{10} r_{10} + \vec{e}_{11} r_{11} + \vec{e}_{12} r_{12} + \vec{e}_{13} r_{13} + \vec{e}_{14} r_{14} + \vec{e}_{15} r_{15} \tag{80}$$

The sedenion differential operator $\diamondsuit_{SWHS}$ and its conjugate operator are defined as,

$$\diamondsuit_{SWHS} = \diamondsuit_{SWHS1} + \diamondsuit_{SWHS2} \quad , \quad \diamondsuit^*_{SWHS} = \diamondsuit^*_{SWHS1} + \diamondsuit^*_{SWHS2} \tag{81}$$

In the sedenion field SW-HS, the definition of the field strength is very different to that of the octonionic field, and thus the succeeding equations and operators should be revised. By analogy with the case of other fields, the sedenion differential operator $\diamondsuit$ needs to be generalized to the new operator ($\mathcal{A}_{S-W}/k^A_{S-W} + \mathcal{A}_{H-S}/k^A_{H-S} + \diamondsuit$). This is because the sedenion field SW-HS includes the hyper-strong field and strong-weak field. Therefore the physical characteristics of the sedenion field SW-HS will be much more complicated than that of the octonionic fields in many aspects.

It can be predicted that field strength of off-diagonal subfields may be symmetry along the diagonal subfield in Table 13. The hyper-strong consists of A and B interactions. For example, the field strength of B-strong subfield may be equal to that of strong-B subfield. And it is easy to find that B interaction is interconnected with strong interaction. The physical features of each subfield in sedenion field SW-HS meet requirement of equations set in Table 14.

In the sedenion field SW-HS, the field potential ($a_0, a_1, a_2, a_3, a_4, a_5, a_6, a_7, a_8, a_9, a_{10}, a_{11}, a_{12}, a_{13}, a_{14}, a_{15}$) is defined as

$$\mathcal{A} = \diamondsuit^* \circ \mathcal{X}$$
$$= a_0 + a_1 \vec{e}_1 + a_2 \vec{e}_2 + a_3 \vec{e}_3 + a_4 \vec{e}_4 + a_5 \vec{e}_5$$

$$+ a_6 \vec{e}_6 + a_7 \vec{e}_7 + a_8 \vec{e}_8 + a_9 \vec{e}_9 + a_{10} \vec{e}_{10}$$
$$+ a_{11} \vec{e}_{11} + a_{12} \vec{e}_{12} + a_{13} \vec{e}_{13} + a_{14} \vec{e}_{14} + a_{15} \vec{e}_{15} \quad (82)$$

where, $k_{rx} \mathcal{X} = k^{S\text{-}W}_{rx} \mathcal{X}_{S\text{-}W} + k^{H\text{-}S}_{rx} \mathcal{X}_{H\text{-}S}$. And $\mathcal{X}_{S\text{-}W}$ and $\mathcal{X}_{H\text{-}S}$ are the physical quantity in the strong-weak field and hyper-strong field respectively.

Table 13. Subfields of sedenion field SW-HS

| | A Interaction | B Interaction | Strong Interaction | Weak Interaction |
|---|---|---|---|---|
| A quaternionic space A | A-A subfield<br>A A-charge<br>$\gamma^B_{aa}$ | B-A subfield<br>A B-charge<br>$\gamma^B_{ba}$ | strong-A subfield<br>A strong-charge<br>$\gamma^B_{sa}$ | weak-A subfield<br>A weak-charge<br>$\gamma^B_{wa}$ |
| B quaternionic space B | A-B subfield<br>B A-charge<br>$\gamma^B_{ab}$ | B-B subfield<br>B B-charge<br>$\gamma^B_{bb}$ | strong-B subfield<br>B strong-charge<br>$\gamma^B_{sb}$ | weak-B subfield<br>B weak-charge<br>$\gamma^B_{wb}$ |
| strong quaternionic space S | A-strong subfield<br>S A-charge<br>$\gamma^B_{as}$ | B-strong subfield<br>S B-charge<br>$\gamma^B_{bs}$ | strong-strong subfield<br>S strong-charge<br>$\gamma^B_{ss}$ | weak-strong subfield,<br>S weak-charge<br>$\gamma^B_{ws}$ |
| weak quaternionic space W | A-weak subfield<br>W A-charge<br>$\gamma^B_{aw}$ | B-weak subfield<br>W B-charge<br>$\gamma^B_{bw}$ | strong-weak subfield<br>W strong-charge<br>$\gamma^B_{sw}$ | weak-weak subfield<br>W weak-charge<br>$\gamma^B_{ww}$ |

The field strength $\mathcal{B}$ of the sedenion field SW-HS can be defined as
$$\mathcal{B} = (\mathcal{A}/K + \lozenge) \circ \mathcal{A} = (\mathcal{A}_{S\text{-}W} / k^A_{S\text{-}W} + \mathcal{A}_{H\text{-}S} / k^A_{H\text{-}S} + \lozenge) \circ \mathcal{A} \quad (83)$$

where, $k^A_{S\text{-}W}$ and $k^A_{H\text{-}S}$ are coefficients. $K = K^A_{SWHS}$ is the coefficient. $\mathcal{A}_{S\text{-}W}$ is the field potential in the strong-weak field, and $\mathcal{A}_{H\text{-}S}$ is the field potential in the hyper-strong field.

Then the field source and the force of the sedenion field can be defined respectively as
$$K \mu \mathcal{S} = K (\mathcal{A}_{S\text{-}W} / k^A_{S\text{-}W} + \mathcal{A}_{H\text{-}S} / k^A_{H\text{-}S} + \lozenge)^* \circ \mathcal{B} \quad (84)$$
$$\mathcal{Z} = K (\mathcal{A}_{S\text{-}W} / k^A_{S\text{-}W} + \mathcal{A}_{H\text{-}S} / k^A_{H\text{-}S} + \lozenge) \circ \mathcal{S} \quad (85)$$

where, the mark (*) denotes sedenion conjugate. The coefficient μ is interaction intensity of the sedenion field HS-SW.

The angular momentum of the sedenion field can be defined as ($k_{rx}$ is the coefficient)
$$\mathcal{M} = \mathcal{S} \circ (\mathcal{R} + k_{rx} \mathcal{X}) \quad (86)$$

and the energy and power in the sedenion field can be defined respectively as
$$\mathcal{W} = K (\mathcal{A}_{S\text{-}W} / k^A_{S\text{-}W} + \mathcal{A}_{H\text{-}S} / k^A_{H\text{-}S} + \lozenge)^* \circ \mathcal{M} \quad (87)$$
$$\mathcal{N} = K (\mathcal{A}_{S\text{-}W} / k^A_{S\text{-}W} + \mathcal{A}_{H\text{-}S} / k^A_{H\text{-}S} + \lozenge) \circ \mathcal{W} \quad (88)$$

Table 14.  Equations set of sedenion field SW-HS

| Spacetime | Sedenion space SW-HS |
|---|---|
| $\mathcal{X}$ physical quantity | $k_{rx}\mathcal{X} = k^{S\text{-}W}_{rx}\mathcal{X}_{S\text{-}W} + k^{H\text{-}S}_{rx}\mathcal{X}_{H\text{-}S}$ |
| Field potential | $\mathcal{A} = \lozenge^* \circ \mathcal{X}$ |
| Field strength | $\mathcal{B} = (\mathcal{A}_{S\text{-}W}/k^A_{S\text{-}W} + \mathcal{A}_{H\text{-}S}/k^A_{H\text{-}S} + \lozenge) \circ \mathcal{A}$ |
| Field source | $K\mu S = K(\mathcal{A}_{S\text{-}W}/k^A_{S\text{-}W} + \mathcal{A}_{H\text{-}S}/k^A_{H\text{-}S} + \lozenge)^* \circ \mathcal{B}$ |
| Force | $Z = K(\mathcal{A}_{S\text{-}W}/k^A_{S\text{-}W} + \mathcal{A}_{H\text{-}S}/k^A_{H\text{-}S} + \lozenge) \circ S$ |
| Angular momentum | $\mathcal{M} = S \circ (\mathcal{R} + k_{rx}\mathcal{X})$ |
| Energy | $\mathcal{W} = K(\mathcal{A}_{S\text{-}W}/k^A_{S\text{-}W} + \mathcal{A}_{H\text{-}S}/k^A_{H\text{-}S} + \lozenge)^* \circ \mathcal{M}$ |
| Power | $\mathcal{N} = K(\mathcal{A}_{S\text{-}W}/k^A_{S\text{-}W} + \mathcal{A}_{H\text{-}S}/k^A_{H\text{-}S} + \lozenge) \circ \mathcal{W}$ |

In sedenion field SW-HS, the Dirac equation and Klein-Gordon equation can be attained respectively from the energy equation (87) and power equation (88) after substituting operator $K(\mathcal{A}_{S\text{-}W}/k^A_{S\text{-}W} + \mathcal{A}_{H\text{-}S}/k^A_{H\text{-}S} + \lozenge)$ for $(\mathcal{W}_{S\text{-}W}/k^A_{S\text{-}W}b^A_{S\text{-}W} + \mathcal{W}_{H\text{-}S}/k^A_{H\text{-}S}b^A_{H\text{-}S} + \lozenge)$. The coefficients $b^A_{S\text{-}W}$ and $b^A_{H\text{-}S}$ are the Plank-like constant.

The $\mathcal{U}$ equation of the quantum mechanics can be defined as

$$\mathcal{U} = (\mathcal{W}_{S\text{-}W}/k^A_{S\text{-}W}b^A_{S\text{-}W} + \mathcal{W}_{H\text{-}S}/k^A_{H\text{-}S}b^A_{H\text{-}S} + \lozenge)^* \circ \mathcal{M} \tag{89}$$

The $\mathcal{L}$ equation of the quantum mechanics can be defined as

$$\mathcal{L} = (\mathcal{W}_{S\text{-}W}/k^A_{S\text{-}W}b^A_{S\text{-}W} + \mathcal{W}_{H\text{-}S}/k^A_{H\text{-}S}b^A_{H\text{-}S} + \lozenge) \circ \mathcal{U} \tag{90}$$

The three sorts of Dirac-like equations can be obtained from the Eqs.(83), (84) and (85) respectively. The $\mathcal{G}$ equation of the quantum mechanics can be defined as

$$\mathcal{G} = (\mathcal{W}_{S\text{-}W}/k^A_{S\text{-}W}b^A_{S\text{-}W} + \mathcal{W}_{H\text{-}S}/k^A_{H\text{-}S}b^A_{H\text{-}S} + \lozenge) \circ \mathcal{A} \tag{91}$$

The $\mathcal{T}$ equation of the quantum mechanics can be defined as

$$\mathcal{T} = (\mathcal{W}_{S\text{-}W}/k^A_{S\text{-}W}b^A_{S\text{-}W} + \mathcal{W}_{H\text{-}S}/k^A_{H\text{-}S}b^A_{H\text{-}S} + \lozenge)^* \circ \mathcal{G} \tag{92}$$

The $O$ equation of the quantum mechanics can be defined as

$$O = (\mathcal{W}_{S\text{-}W}/k^A_{S\text{-}W}b^A_{S\text{-}W} + \mathcal{W}_{H\text{-}S}/k^A_{H\text{-}S}b^A_{H\text{-}S} + \lozenge) \circ \mathcal{T} \tag{93}$$

Table 15.  Quantum equations set of sedenion field SW-HS

| Energy quantum | $\mathcal{U} = (\mathcal{W}_{S\text{-}W}/k^A_{S\text{-}W}b^A_{S\text{-}W} + \mathcal{W}_{H\text{-}S}/k^A_{H\text{-}S}b^A_{H\text{-}S} + \lozenge)^* \circ \mathcal{M}$ |
|---|---|
| Power quantum | $\mathcal{L} = (\mathcal{W}_{S\text{-}W}/k^A_{S\text{-}W}b^A_{S\text{-}W} + \mathcal{W}_{H\text{-}S}/k^A_{H\text{-}S}b^A_{H\text{-}S} + \lozenge) \circ \mathcal{U}$ |
| Field strength quantum | $\mathcal{G} = (\mathcal{W}_{S\text{-}W}/k^A_{S\text{-}W}b^A_{S\text{-}W} + \mathcal{W}_{H\text{-}S}/k^A_{H\text{-}S}b^A_{H\text{-}S} + \lozenge) \circ \mathcal{A}$ |
| Field source quantum | $\mathcal{T} = (\mathcal{W}_{S\text{-}W}/k^A_{S\text{-}W}b^A_{S\text{-}W} + \mathcal{W}_{H\text{-}S}/k^A_{H\text{-}S}b^A_{H\text{-}S} + \lozenge)^* \circ \mathcal{G}$ |
| Force quantum | $O = (\mathcal{W}_{S\text{-}W}/k^A_{S\text{-}W}b^A_{S\text{-}W} + \mathcal{W}_{H\text{-}S}/k^A_{H\text{-}S}b^A_{H\text{-}S} + \lozenge) \circ \mathcal{T}$ |

In the sedenion field SW-HS, the intermediate and field source particles can be achieved. We can find the intermediate particles and other kinds of new and unknown particles which may be existed in the nature.

If the preceding sedenion fields are extended to other sorts of sedenion fields in Table 2, many extra and also more complicated equations set will be gained in the similar way.

## 7. Conclusions

By analogy with the four sorts of octonionic fields, the twelve sorts of sedenion fields have been developed, including their field equations, quantum equations and some new unknown particles.

In the sedenion field SW-EG, the study deduces the Yang-Mills equation, Dirac equation, Schrodinger equation and Klein-Gordon equation of the quarks and leptons etc. It infers the a few Dirac-like equations of intermediate particles among quarks and leptons. It draws some conclusions of field source particles and intermediate particles which are consistent with current electro-weak theory. It predicts that there are some new unknown particles of field sources (quarks and leptons) and their intermediate particles in the sedenion field SW-EG.

In the sedenion field HS-SW, the study deduces the Dirac equation, Schrodinger equation, Klein-Gordon equation and some newfound equations of the sub-quarks etc. It infers the Dirac-like equations of intermediate particles among sub-quarks etc. It predicts that there are some new unknown particles of field sources (sub-quarks etc.) and their intermediate particles in the sedenion field HS-SW.

In the sedenion field theory, we can find that the interplays among the different interactions are much more complicated than we found and thought before.

## Acknowledgements

The author thanks Minfeng Wang, Yun Zhu, Zhimin Chen, Shaohan Lin and Xu Chen for helpful discussions. This project was supported by National Natural Science Foundation of China under grant number 60677039, Science & Technology Department of Fujian Province of China under grant number 2005HZ1020 and 2006H0092, and Xiamen Science & Technology Bureau of China under grant number 3502Z20055011.

## References

[1] Zihua Weng. Octonionic electromagnetic and gravitational interactions and dark matter. arXiv: physics /0612102.
[2] Zihua Weng. Octonionic quantum interplays of dark matter and ordinary matter. arXiv: physics /0702019.
[3] A. Denner, B. Jantzen, S. Pozzorini. Two-loop electroweak next-to-leading logarithmic corrections to massless fermionic processes. Nucl. Phys. B, 761 (2007) 1-62.
[4] Zihua Weng. Octonionic strong and weak interactions and their quantum equations. arXiv: physics /0702054.
[5] Zihua Weng. Octonionic hyper-strong and hyper-weak fields and their quantum equations. arXiv: physics /0702086.